\documentclass[superscriptaddress, nobalancelastpage,
twocolumn,prl]{revtex4-1}
\pdfoutput=1
\usepackage[normalem]{ulem}
\usepackage{cancel}
\usepackage{graphicx}
\usepackage{amssymb}
\usepackage{mathtools}
\usepackage{mathrsfs}
\usepackage{graphicx,psfrag,xspace}
\usepackage{amsmath}
\usepackage{braket}
\usepackage{setspace}
\usepackage[usenames,dvipsnames,svgnames,table]{xcolor}
\usepackage{color}
\usepackage[colorlinks,bookmarks=false,citecolor=NavyBlue,linkcolor=OliveGreen,urlcolor=blue]{hyperref}
\usepackage{comment}

\def\pp{\upsilon}

\def\gr{\varphi}

\newcommand{\be}{\begin{equation}}
\newcommand{\ee}{\end{equation}}
\newcommand{\ba}{\begin{aligned}}
\newcommand{\ea}{\end{aligned}}

\def\doi{http://dx.doi.org/}

\newcommand{\titleinfo}{Analytic solution of the Domain Wall non-equilibrium
stationary state} 

\date{\today}

\begin{document}
\title{\titleinfo} 
\author{Mario Collura}
\author{Andrea De Luca}
\affiliation{The Rudolf Peierls Centre for Theoretical Physics, Oxford University, 
Oxford, OX1 3NP, United Kingdom.}
\author{Jacopo Viti}
\affiliation{ECT \& Instituto Internacional de Fisica, UFRN, Lagoa Nova 59078-970 Natal, Brazil}

\begin{abstract}
We consider the out-of-equilibrium dynamics generated by joining 
two domains with arbitrary opposite magnetisations. We study the stationary state which emerges
by the unitary evolution via the spin $1/2$ XXZ Hamiltonian, in the gapless regime, where the system 
develops a stationary spin current. Using the generalized
hydrodynamic approach, we present a simple formula for the space-time profile of the 
spin current and the magnetisation exact in the limit of large times. As a remarkable effect, we show that the
stationary state has a strongly discontinuous dependence on the strength of interaction. This feature allows us 
to give a qualitative estimation for the transient behavior of the current which is compared with numerical simulations. 
Moreover, we analyse the behavior around the edge of the magnetisation profile and we argue that, unlike the XX free-fermionic point, 
interactions always prevent the emergence of a Tracy-Widom scaling. 
\end{abstract}


\maketitle

\paragraph{Introduction.---}
Recent experimental developments with cold atoms~\cite{bdz-08} have given a new
perspective to the study of non-equilibrium transport under coherent evolution. As an example, the measurement of conductances well beyond the regime of linear response has provided clear examples
of the thermoelectric effect~\cite{brantut-12, brantut-13}.
The simplest protocol to induce an out-of-equilibrium behavior is the one of quantum quenches, in which the system is prepared in an equilibrium state of the initial Hamiltonian ${\boldsymbol H}_0$,
which is suddenly switched to ${\boldsymbol H}$, thus inducing a non-trivial time-evolution~\cite{cc-06, pssv-11,
ccgo-11, CaEM16}. Then, in describing the long-time dynamics,  a fundamental role is played by the conserved quantities of $\boldsymbol{H}$, i.e. the set of local (or quasi-local~\cite{prosenreview}) operators $\{\boldsymbol{Q}_k\}$ satisfying $[\boldsymbol{Q}_k, \boldsymbol{H}]=0$. As the system is isolated, the expectation value of these conserved quantities remains constant during the  evolution. For homogeneous systems, these conditions are sufficient to predict the exact behavior of any local observable at long times: this is based on assuming equilibration to the \textit{generalized Gibbs ensemble} (GGE), which results from the maximization of entropy given the constraints imposed by conserved quantities~\cite{ViRi16}. 
This principle suggests a dichotomy between generic models, for which a finite number 
of conserved quantities exist (i.e. the Hamiltonian and few others) and integrable ones,
which instead present an infinite number of them~\cite{booksIntegrability}. Nowadays,
the GGE scheme has been validated by 
several 
experiments \cite{kww-06, HLFS07, gklk-12, fse-13, lgkr-13, glms-14, langen-15, LaGS16} and
theoretical works, employing free theories~\cite{fagottiIsingPRL, fagottiIsingJSTAT, EsFa16
}, integrability
\cite{ IDWC15,
IlievskiPRL,
IlievskiJSTAT,
PiVC16,
delu2016, MBPC17
} and numerical methods~\cite{rdyo-07, brand2012, ViRi16, ViIR17}.

However, the study of transport requires considering more generic situations, where
for instance, an initial spatial inhomogeneity is used to induce particle or energy flow. 
The simplest examples are \textit{local quenches} where $\boldsymbol{H}_0$ and $\boldsymbol{H}$
differ only in a finite region of space, due, for instance, to the presence of a localized defect~\cite{CaCa07, JMSJD11, delucaFlux, F16, BeFa16, DeLucaAlvise, PBPD17}.
In particular, in the \textit{partitioned protocol}, the initial density matrix is factorized into two halves, 
i.e. $\rho_0 = \rho_L \otimes \rho_R$, which are suddenly connected, inducing an out-of-equilibrium dynamics around the junction~\cite{BeDo12,BeDo15}.
This problem was well-understood in non-interacting theories~\cite{
ARRS99,
PlKa07,
EiRz13,
DVBD13,
CoKa14,
EiZi14,
DeMV15,
VSDH16,
ADSV16,
KoZi16,
Korm17,
PeGa17} (with even mathematically rigorous treatments \cite{AsPi03,AsBa06}) and field theories~\cite{BeDo12,
BeDo15,
BeDo16,
BeDo16Review,
SoCa08,
CaHD08,
Mint11,
MiSo13,
DoHB14,
LLMM17,
SDCV17,
DuSC17}, even in higher dimensions~\cite{CoMa14,BDLSNature, DLSB15}.
However, for interacting models, only numerical approaches~\cite{GKSS05, LaMi10, 
KaBM12, KaMH14, SaMi13,AlHe14,BDVR16,
SHZB16, Karr17} 
and approximate results were available~\cite{DVMR14,CCDH14,Doyo15,VaKM15,Zoto16}. 
While at 
extremely long times $v_{0} t \gg L$ (with $L$ the system length and $v_0$ the maximal
velocity~\cite{LR72, bonnes14, Bert17}, one expects the system to become homogeneous, 
the most interesting regime is the one where $a \ll v_0 t \ll L$ (with $a$ the typical microscopic length).
In this regime, conserved quantities are dynamically exchanged between different portions of space
and therefore the simple knowledge of their initial value is not enough to characterize the local behavior of the steady state. 
Nevertheless, conserved quantities must still satisfy a continuity equation $\partial_t \boldsymbol{q}_k (x,t) + \partial_t \boldsymbol{j}_k(x,t) = 0$, with $\boldsymbol{q}_k(x,t)$ the local density 
of $\boldsymbol{Q}_k$ and $\boldsymbol{j}_k(x,t)$ the corresponding current. This condition was recently employed \cite{CaDY16, BCDF16} to derive a \textit{generalized hydrodynamic description} (GHD) applicable to a large class of one-dimensional integrable models \cite{DeLuca16,DoYo16,DoSY17,DoSp17,DoYC17,DDKY17,BVKM17, IlDe17, DoSp17-2, PDCBF17, IlDeHubbard}. 
For the partitioned protocol, this description becomes exact and at large times, 
a \textit{Local quasi-stationary state} (LQSS) emerges, in which
local observables only depend on the scaling variable $\zeta = x/t$.

In this letter, we consider the XXZ spin $1/2$ in the gapless regime, prepared in the partitioned initial state composed by two domains of arbitrary opposite magnetisations. We solve the hydrodynamic equations, obtaining simple analytic expressions for the
magnetisation and spin current profiles. 
To the best of our knowledge, this represents a unique example of an out-of-equilibrium steady state of an interacting quantum system, admitting an explicit exact solution. Remarkably, 
the stationary spin current exhibits a strongly discontinuous behavior as a function of the anisotropy, as a result of the peculiar structure of bound states in the model. 


\paragraph{The Model.---}
We consider the XXZ Hamiltonian
\be
{\boldsymbol H}= \sum_{i=-\frac{L}{2}}^{\frac{L}{2}-1}\left[{\boldsymbol s}^{x}_i {\boldsymbol s}^{x}_{i+1}+  {\boldsymbol s}^{y}_i {\boldsymbol s}^{y}_{i+1}+\Delta \left( {\boldsymbol s}^{z}_i {\boldsymbol s}^{z}_{i +1}-\frac{1}{4}\right)\right]\;,
\label{eq:H}
\ee
where $L$ is the length of the chain and
${\boldsymbol s}^{\alpha}_{i}$ are spin-$1/2$ operators 
each acting on the local Hilbert space at site $i$. 
We focus on the gapless phase, thus specializing   
$\Delta = \cos(\gamma)$ with $\gamma = \pi\,Q/P$, where $Q$ and $P$ are 
two coprime integers with $1\le Q<P$. The ratio $Q/P$ admits a finite continued fraction representation
$Q/P=[0;\nu_1,\nu_2,\dots,\nu_{\delta}]$ with length $\delta$.
For any finite $L$ such model is exactly solvable via Bethe-ansatz method~\cite{takahashi, gaudin, korepin}.
In the thermodynamic limit (i.e. when $L\to\infty$ with fixed particle density), 
a generic thermodynamic state can be fully characterized by a set of 
functions $\{\rho_{j}(\lambda), \rho^{h}_{j}(\lambda)\}$ with
$j\in\{1,\dots, \ell \}$, $\ell = \sum_{j=1}^{\delta} \nu_{j}$ and $\lambda \in [-\infty,\infty]$.
These functions, also known as ``root densities'', describe different species of quasiparticles (different ``strings'')
and are solutions of a system of coupled nonlinear integral equations~\cite{takahashi, gaudin, korepin}.
We can associate to each string $j$ a given {\it parity} $\pp_j \in \{-1,1\}$, {\it length} 
$n_j \in \{ 1, \dots, P-1\}$ and  {\it sign} $\sigma_j \in \{-1,1\}$.
The \textit{filling factors} are introduced as the ratios $\vartheta_{j}(\lambda) \equiv \rho_{j}(\lambda)/[\rho_{j}(\lambda)+\rho^{h}_{j}(\lambda)]$; we refer the reader to the 
Supplemental Material \cite{supplmat} for further details.

\paragraph{The Local Quasi Stationary State Redux.---}
Starting from a partitioned initial state 
$\varrho_0 = \varrho_L \otimes \varrho_R$, and unitarily evolving this state under the 
Hamiltonian (\ref{eq:H}), the general formal solution of 
the LQSS on the space-time coordinates $x,\,t$ reads~\cite{BCDF16}
\be\label{eq:LQSS}
\vartheta_{j, \zeta}(\lambda) = 
\vartheta_j^L(\lambda) \Theta(v_{j,\zeta} (\lambda) - \zeta ) + 
\vartheta_j^R(\lambda) \Theta(\zeta - v_{j,\zeta} (\lambda) ) \;, 
\ee
in terms of the scaling variable $\zeta = x/t$, with $\Theta(z)$ being the Heaviside step function.  

The functions $\vartheta^{L/R}_j(\lambda)$ 
are the filling factors which describe the homogeneous stationary state 
emerging on the very far left/right part of the system.
Eq. (\ref{eq:LQSS}) formally identifies, for each type of quasiparticles, 
the related stationary distribution function. This solution admits a very simple geometrical interpretation:
for any value of $j$ and $\lambda$, starting from $\zeta = - \infty$, the left bulk stationary description 
extends up to $\zeta^{*}_{j}(\lambda)$, such that $v_{j,\zeta^*}(\lambda) =\zeta^*$,
thereafter, $\vartheta_{j,\zeta}(\lambda)$ suddenly jumps to the right bulk stationary description.
In practice this formal solution explicitly depends on the dressed velocity
\be\label{eq:v_dressed}
v_{j,\zeta}(\lambda) = \frac{e'_{j,\zeta}(\lambda)}{p'_{j,\zeta}(\lambda)}\; ,
\ee
where $e'_{j,\zeta}(\lambda)$ and $p'_{j,\zeta}(\lambda)$ are respectively the dressed 
energy and momentum derivative.
For a generic thermodynamic state described by a set of filling factors $\{\vartheta_j(\lambda)\}$, and a generic
conserved charge ${\boldsymbol Q}$ with single particle eigenvalues $\mathfrak{q}_j(\lambda)$,
the dressing is obtained solving
\be\label{eq:dress_q}
q'_j(\lambda) = \mathfrak{q}'_j(\lambda) 
- \sum_k \int d\mu \,T_{j,k}(\lambda - \mu) \sigma_k \vartheta_k(\mu) q'_k(\mu)\; ,
\ee
where we chose the convention to use 
calligraphic notation for bare quantities $\mathfrak{q}_j(\lambda)$.
Introducing the function
\be
a_{n}^{(\pp)}(\lambda) = \frac{\pp}{\pi} \frac{\sin(\gamma n)}{\cosh(2\lambda) - \pp \cos(\gamma n)} \; ;
\ee
the kernel $T_{j,k}(\lambda)$ assumes the form
\begin{align}
T_{j,k}(\lambda) &= (1- \delta_{n_j, n_k}) a_{|n_j - n_k|}^{(\pp_j \pp_k)}(\lambda) 
+ 2 a_{|n_j - n_k|+2}^{(\pp_j \pp_k)}(\lambda) \nonumber \\
&+ \ldots + 2 a_{n_j + n_k - 2}^{(\pp_j \pp_k)}(\lambda)  + a_{n_j + n_k}^{(\pp_j \pp_k)}(\lambda)
\end{align}
while the bare eigenvalues for the energy and the momentum derivative are
\be
\mathfrak{e}_j(\lambda) = -\pi \sin(\gamma)  a_j(\lambda) \;, \quad 
\mathfrak{p}_j'(\lambda) = 2 \pi a_j(\lambda) 
\ee
where we defined $a_{j}(\lambda) \equiv a_{n_j}^{(\pp_j)}(\lambda)$.
Note that, as the dressing operation \eqref{eq:dress_q}
is performed over the state $\vartheta_j(\lambda) = \vartheta_{j,\zeta}(\lambda)$,
the solution for the LQSS has to be found self consistently in such a way that 
it keeps the form in Eq. (\ref{eq:LQSS}) with its own dressed velocity \eqref{eq:v_dressed}.
Therefore, in general, the dressed velocity will depend 
on the scaling variable $\zeta$, via the state $\vartheta_{j,\zeta}(\lambda)$.

From the thermodynamic Bethe ansatz (TBA) description of the local quasi stationary state
we can easily evaluate the expectation value of a generic charge density 
${\boldsymbol q} = {\boldsymbol Q}/L$
\be
\langle {\boldsymbol q}\rangle_{\zeta} 
= \sum_{k} \int \frac{d\lambda}{2\pi} \,  \mathfrak{q}_{k}(\lambda) 
 \sigma_{k} p'_{k,\zeta}(\lambda)  \vartheta_{k,\zeta}(\lambda)\;,
\ee
and the associated current density
\be
\langle {\boldsymbol j}_{\boldsymbol q}\rangle_{\zeta} 
= \sum_{k} \int \frac{d\lambda}{2\pi} \,  \mathfrak{q}_{k}(\lambda) v_{k,\zeta}(\lambda)
 \sigma_{k} p'_{k,\zeta}(\lambda)  \vartheta_{k,\zeta}(\lambda)\; .
\ee


\paragraph{Opposite magnetisation domains.---}
The system is initially prepared into two halves  
with infinite temperature and opposite
values of magnetic field $h$ in the $\hat z$ direction, namely 
\be\label{eq:rho0_mu}
\varrho_0 \equiv \varrho_L(h) \otimes \varrho_R(-h) =
\frac{e^{2 h {\boldsymbol S}^{z}_{L}}}{Z_L}
\otimes
\frac{e^{-2 h  {\boldsymbol S}^{z}_{R}}}{Z_R} \; ,
\ee
where $ {\boldsymbol S}^{z}_{L/R} = \sum_{i \in L/R } {\boldsymbol s}^{z}_{i}$ 
is the $\hat z$-component of the total spin in the left/right part of the system.

A generic thermodynamic state $\varrho_{L/R}(h)$ is stationary 
under the unitary evolution induced by its own XXZ Hamiltonian.
It admits a TBA description in terms of 
constant filling factors $\vartheta^{(h)}_{j}$ (i.e. independent of the rapidity $\lambda$),
which satisfy the major properties (see \cite{supplmat} for the complete definition of $\vartheta^{(h)}_{j}$, $\forall\, j\in \{1,\dots,\ell\}$)
\be
\label{eq:symm}
\vartheta^{(h)}_{j} = \vartheta^{(-h)}_{j} \;\;  j<\ell-1, \quad
\vartheta^{(h)}_{\ell-1} = 1 - \vartheta^{(-h)}_{\ell} \;.
\ee
%
%
In the limit $h\to \infty$ the state $\varrho_{0}$ reduces to the Domain Wall (DW) 
$\ket{\Uparrow}\otimes\ket{\Downarrow}$ product state, with
$\vartheta^{\ket{ \Uparrow } }_{j} = 0$ and $\vartheta^{ \ket{ \Downarrow } }_{j} = \delta_{j,\ell} +  \delta_{j,\ell-1}$, for $j=1,\ldots, \ell$.

\paragraph{The full analytic solution.---}
Now if we consider the protocol generated attaching 
two states with $h$ (left) and $-h$ (right), the $\zeta \to \pm \infty$ boundary conditions
in Eq. (\ref{eq:LQSS}) read $\vartheta^{L}_{j}(\lambda) = \vartheta^{(h)}_{j}$, 
$\vartheta^{R}_{j}(\lambda) = \vartheta^{(-h)}_{j}$. Thanks to the symmetries \eqref{eq:symm} of the boundary filling factors, when constructing the LQSS, only the filling factors  $\vartheta_{j, \zeta}(\lambda)$ corresponding to the  last two strings $j=\ell-1$ and $j=\ell$ may depend on $\zeta$.
In order to fix them, we need to determine the dressed velocities. Using that
 $T_{\ell,k}(\lambda) = -T_{\ell-1, k}(\lambda)$ and $a_{\ell}(\lambda) = - a_{\ell-1}(\lambda)$,
we have
\be\label{eq:p_sym}
p'_{\ell,\zeta}(\lambda) = - p'_{\ell-1,\zeta}(\lambda) \; , \quad
e'_{\ell,\zeta}(\lambda) = - e'_{\ell-1,\zeta}(\lambda)\; ,
\ee
which implies
$v_{\ell,\zeta}(\lambda) = v_{\ell-1,\zeta}(\lambda)$.
As the last two strings always have opposite {\it sign}, i.e. $\sigma_{\ell-1} = -\sigma_{\ell}$,
we can reduce Eq.~\eqref{eq:dress_q} for the dressed momentum derivative to 
\begin{align*}
& p'_{j,\zeta}(\lambda)  = \mathfrak{p}'_{j}(\lambda) 
- \sum_{k\le \ell-2} \sigma_k \vartheta^{(h)}_{k}  \int d\mu \,T_{j,k}(\lambda - \mu) p'_{k,\zeta}(\mu) \nonumber \\
& - \sigma_{\ell}  \int d\mu \, [\vartheta_{\ell,\zeta}(\mu) - \vartheta_{\ell-1,\zeta}(\mu)]  T_{j,\ell}(\lambda - \mu) p'_{\ell,\zeta}(\mu), 
\end{align*}
which does not depend on the space-time scaling variable $\zeta$, since
$\vartheta_{\ell,\zeta}(\mu) - \vartheta_{\ell-1,\zeta}(\mu) = \vartheta^{(h)}_{\ell} - \vartheta^{(h)}_{\ell-1} $.
From now on we discard the subscript $\zeta$ whenever it will be superfluous. 
As a consequence of the last result, we can calculate the dressed momentum derivative 
solving 
\be\label{eq:dress_h}
p'_{j}(\lambda)  = \mathfrak{p}'_{j}(\lambda) 
- \sum_{k} \sigma_k \vartheta^{(h)}_{k}  \int d\mu \,T_{j,k}(\lambda - \mu) p'_{k}(\mu),
\ee
which correspond to evaluate the dressing on the left thermodynamic state $\varrho_{L}(h)$. 
Note that the dressing can be equivalently evaluated in the right part of the system, as
it is even in sign of the magnetic field. 
Eq. (\ref{eq:dress_h}) can be solved in Fourier transform, reducing to an algebraic system of linear equations.
For the last two strings the dressing operation reduces to a simple rescaling
of the bare quantities, i.e.
\be
\label{eq:solution_pprime}
p'_{\ell}(\lambda)  =  {\mathcal R}(h) \, \mathfrak{p}'_{\ell}(\lambda) \; , \quad
p'_{\ell-1}(\lambda)  =  {\mathcal R}(h) \, \mathfrak{p}'_{\ell-1}(\lambda) \; ,
\ee
with the following rescaling factor
\begin{align}
{\mathcal R}(h) & \equiv 
 \frac{\tanh(h)}{2}  \frac{\sinh((n_{\ell} + n_{\ell-1}) h)}{\sinh( n_{\ell} h )\sinh( n_{\ell-1} h )} 
\end{align}
where in the last line we used the relation $n_{\ell}+n_{\ell-1} = P$, and
as expected $\mathcal{R}(-h) = \mathcal{R}(h)$.
As a consequence, the quasiparticle velocity
of the last two strings is not changed by the dressing operation. It can therefore 
be expressed in terms of 
the undressed momentum as follow 
\be
\mathfrak{v}_{\ell} = 
 \frac{\pp_{\ell}\sin(\gamma)}{\sin{(n_{\ell}\gamma})} \sin(\mathfrak{p_{\ell}})
 = \zeta_0 \sin(\mathfrak{\sigma_{\ell} p_{\ell}}) \; ,
\ee
with $\zeta_{0} \equiv \sin(\gamma)/\sin(\pi/P)$ and 
$\sigma_{\ell} \mathfrak{p}_{\ell}(\lambda)$ a strictly increasing function in $[ -\pi/P, \pi/P]$.
Therefore, the velocity 
$\mathfrak{v}_{\ell}(\lambda) \in [ -\sin(\gamma), \sin(\gamma)]$.
The explicit form of the LQSS for the last two strings thus reads (for $j \in \{ \ell-1, \ell \}$)
\be
\vartheta_{j,\zeta}(\lambda)  =
\vartheta^{(h)}_{j} \Theta(\sigma_{j}\mathfrak{p}_{j}-\mathfrak{p}^{*}_{\zeta}) 
+ \vartheta^{(-h)}_{j}\Theta(\mathfrak{p}^{*}_{\zeta}-\sigma_{j}\mathfrak{p}_{j}) \; ,
\ee
where $\mathfrak{p}^{*}_{\zeta} \equiv \arcsin [ \zeta/\zeta_{0} ]$.
From this, using ${\rm Tr}[ {\boldsymbol s}^{z} \varrho_{L}(h)] = \tanh(h)/2$ and 
$\mathcal{R}(h) = \tanh(h)/(1- \vartheta^{(h)}_{\ell}-\vartheta^{(h)}_{\ell-1})$,
we can easily evaluate the magnetisation and spin current profile inside the light-cone $\zeta \in [-\sin(\gamma), \sin(\gamma)]$,
\begin{subequations}
\label{eq:profiles}
\begin{align}
\label{eq:lqss_s}
\langle {\boldsymbol s}^{z}\rangle_{\zeta} 
& = - \frac{\tanh(h)}{2\pi/P} \arcsin \left( \frac{\zeta}{ \zeta_0}  \right)\; ,\\
\label{eq:lqss_js}
\langle {\boldsymbol j}_{{\boldsymbol s}^{z}}\rangle_{\zeta} 
& = \frac{\tanh(h)}{2\pi/P} \zeta_0 
 \left [ \sqrt{1-  \frac {\zeta^2} {\zeta^2_0}} 
- \cos \left(\frac{\pi}{P} \right) \right] \; ,
\end{align}
\end{subequations}
which are simply related one another via 
the continuity equation 
$
\zeta\partial_{\zeta}\langle {\boldsymbol s}^{z}\rangle_{\zeta}
=
\partial_{\zeta}\langle {\boldsymbol j}_{{\boldsymbol s}^{z}}\rangle_{\zeta} \; .
$
Interestingly, the way in which the magnetic field $h$ 
enters in the stationary solutions is almost trivial: 
indeed, Eqs.~\eqref{eq:profiles}
coincide with the DW solutions ($h\to\infty$) simply rescaled by the factor $\tanh(h)$.
Moreover in this limit $\mathcal{R}(h)\to 1$, showing that for the DW initial state, no dressing occurs.
\begin{figure}[t!]
\includegraphics[width=0.48\textwidth]{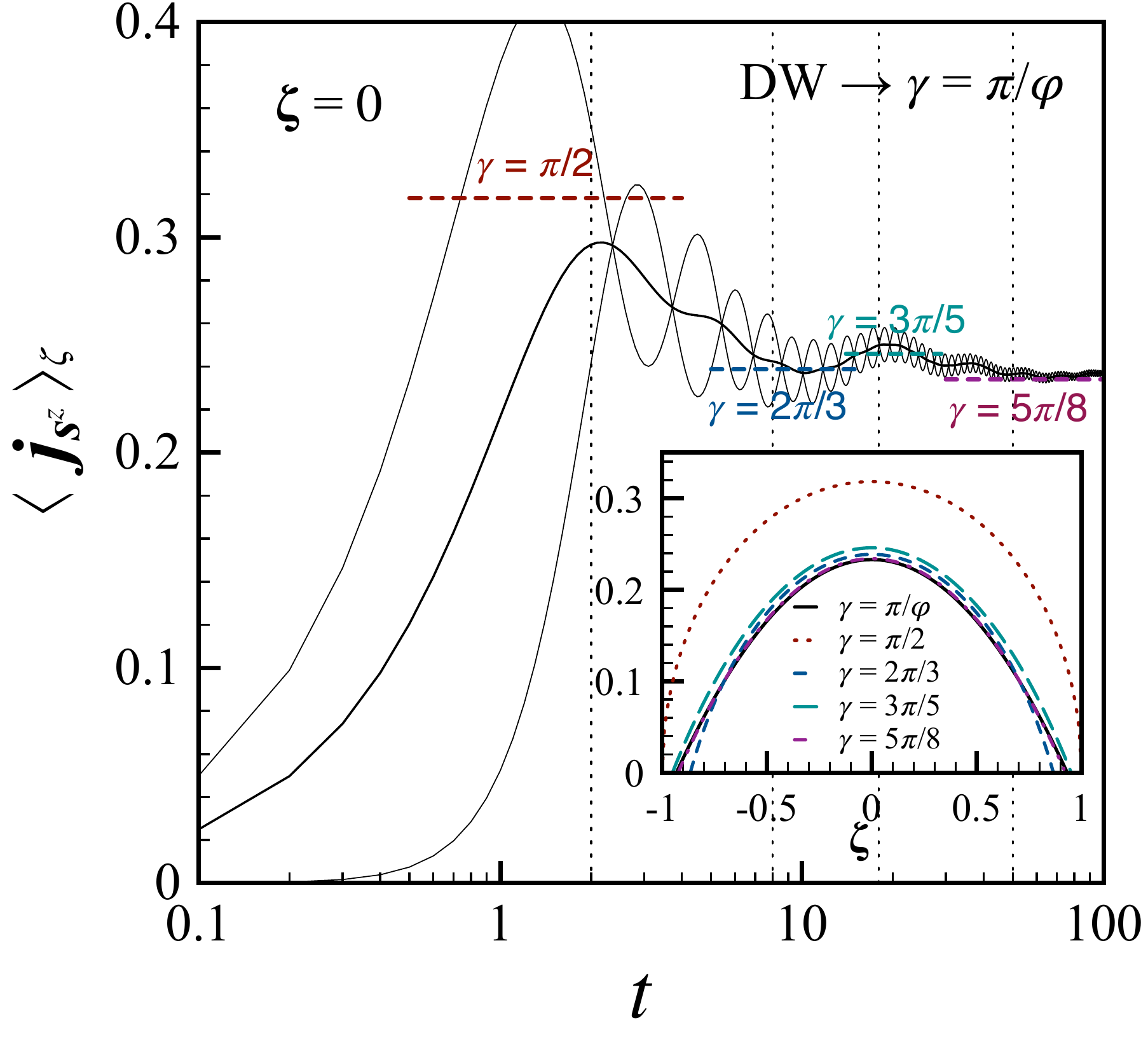}
\caption{\label{fig:goldenratio}
({\bf Main}) Spin current at the junction, i.e. $\langle {\boldsymbol j}_{{\boldsymbol s}^{z}}\rangle_{\zeta=0}$,
for a quench from the DW with $\gamma = \pi/\varphi$. 
The time-dependent DMRG data for the spin current between lattice sites $(0,1)$ and $(1,2)$ (thin black lines)
and their average (thick black line) are compared with the
stationary values associated to the rational approximation of the golden ratio (horizontal dashed lines).
The dotted vertical lines represent the typical timescale at which the current passes from 
one rational approximation to the next one.
({\bf Inset}) The analytic stationary profile for the golden ratio is compared with
different rational approximations.
}
\end{figure}

\paragraph{The anisotropy dependence.---}
It is interesting to investigate  
how the  interaction strength $\Delta$ affects the stationary state. 
Both current and magnetisation profiles have an explicit dependence on the denominator $P$ of $\pi/\gamma$: as one can pick two arbitrarily close values $\gamma = \pi Q/P$ and $\tilde\gamma  = \pi \tilde Q/\tilde P$, with very different values of $P$ and $\tilde P$,
the magnetisation and current profiles exhibit jumps in correspondence of
any rational $\pi/\gamma$, corresponding to a dense subset of $\Delta \in [-1,1]$. 
Nevertheless, the continuation to irrational values 
is well defined taking $P\to \infty$ with $\gamma$ finite. In such limit, the
current profile reduces to (for $\gamma/\pi \in \mathbb{R}/\mathbb{Q}$)
\begin{align}
\label{eq:irrational}
\langle {\boldsymbol j}_{{\boldsymbol s}^{z}}\rangle^{(\mathbb{R}/\mathbb{Q})}_{\zeta}  & 
= \frac{ \tanh(h)}{4} 
\left[ \sin(\gamma) - \frac{\zeta^2}{\sin(\gamma)} \right] \; ,
\end{align}
%
and the magnetisation behaves linearly in $\zeta$.
For any irrational number $\gamma/\pi$, although the large time limit will be characterised by 
the stationary values in (\ref{eq:irrational}), we expect the relaxation dynamics 
to spend long times on the rational approximations of such an irrational, i.e.
the truncated continued fractions $[0; \nu_1,\ldots,\nu_{n}]$.
The ideal case to verify this hypothesis
corresponds to all $\nu_k = 1$,
i.e. $\gamma = \pi/\varphi$, with $\gr \equiv (1 + \sqrt{5})/2$, the {\it golden ratio}.
Its $n$-th order rational approximation is given by $F_{n}/F_{n+1}$, where
$F_{n}$ are the Fibonacci numbers and $1/\gr = \lim_{n\to\infty} F_{n}/F_{n+1} $.
In Fig.~\ref{fig:goldenratio}, the numerical data for the spin current clearly oscillate in time 
between different stationary values associated to different orders of approximation of the golden ratio.
The curve remains close to the $n$-th rational approximation for an exponentially long time, 
$t \propto F_{n}^{2} \simeq \varphi^{2n}$.

Remarkably, our exact result definitively gives analytical confirmation 
to the tightness of the bound in \cite{PrIl13} for the spin Drude weight $\mathcal{D}_{{\boldsymbol s}^{z}}$, numerically corroborated in \cite{IlDe17}.
In the linear response regime indeed, the spin Drude weight gives the magnitude of the singular part of the
spin conductivity, therefore signaling ballistic transport~\cite{CaZP95,HHCB03,SiPA09,Pros11,LHMH11,Znid11}. 
Following \cite{IlDe17}, we integrate the current \eqref{eq:lqss_js} over $\zeta$ to obtain for $\beta \to 0$:
 \begin{align}\label{eq:drude}
 (16/\beta)\mathcal{D}_{{\boldsymbol s}^{z}} = 
  \zeta^{2}_{0} \left[ 1- \frac{\sin(2\pi/P)}{2\pi/P} \right]\; .
 \end{align}
This result exactly coincides with the lower bound obtained in \cite{PrIl13},
confirming that it is in fact saturated.

\begin{figure}[t!]
\includegraphics[width=0.24\textwidth]{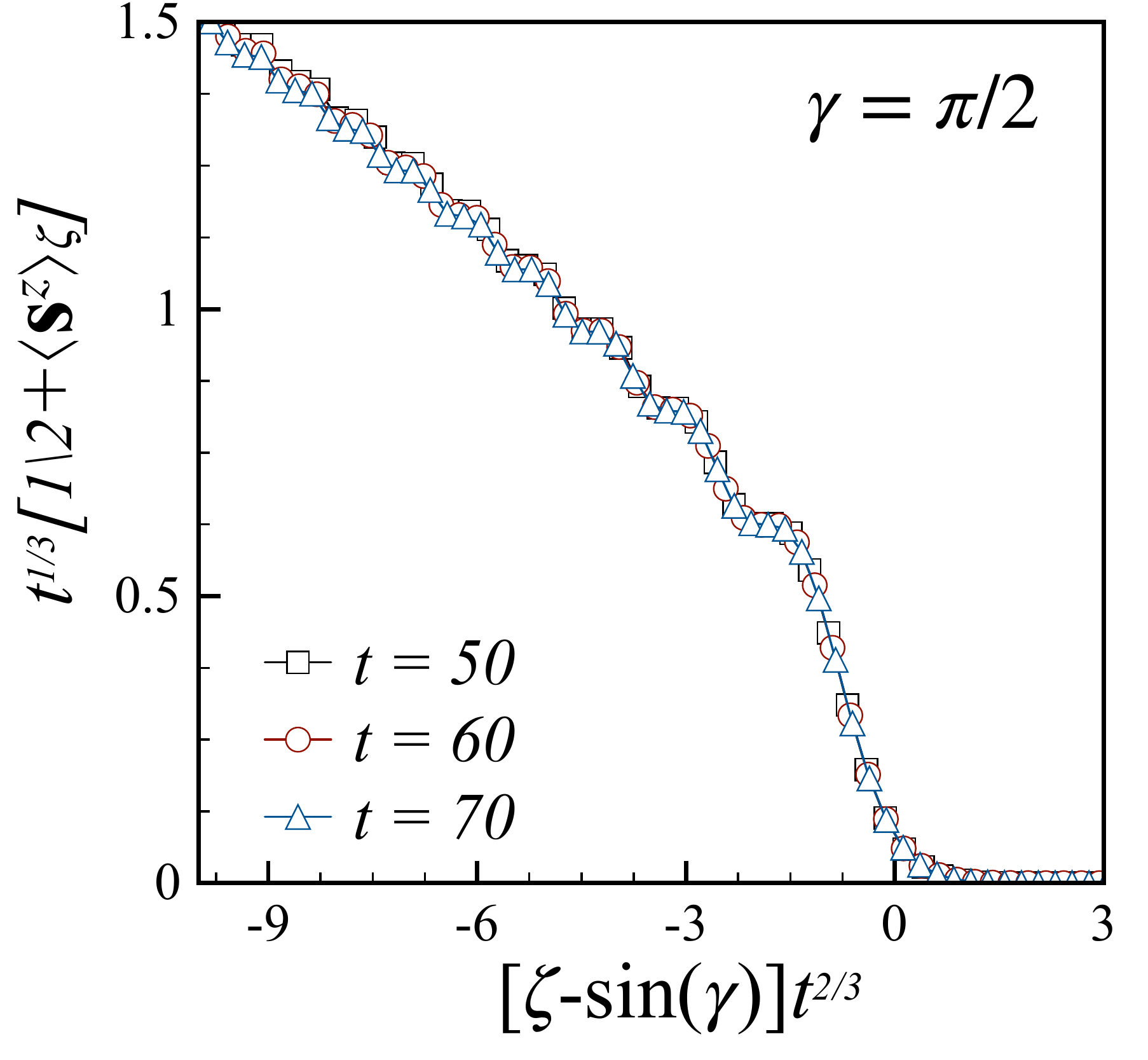}\includegraphics[width=0.24\textwidth]{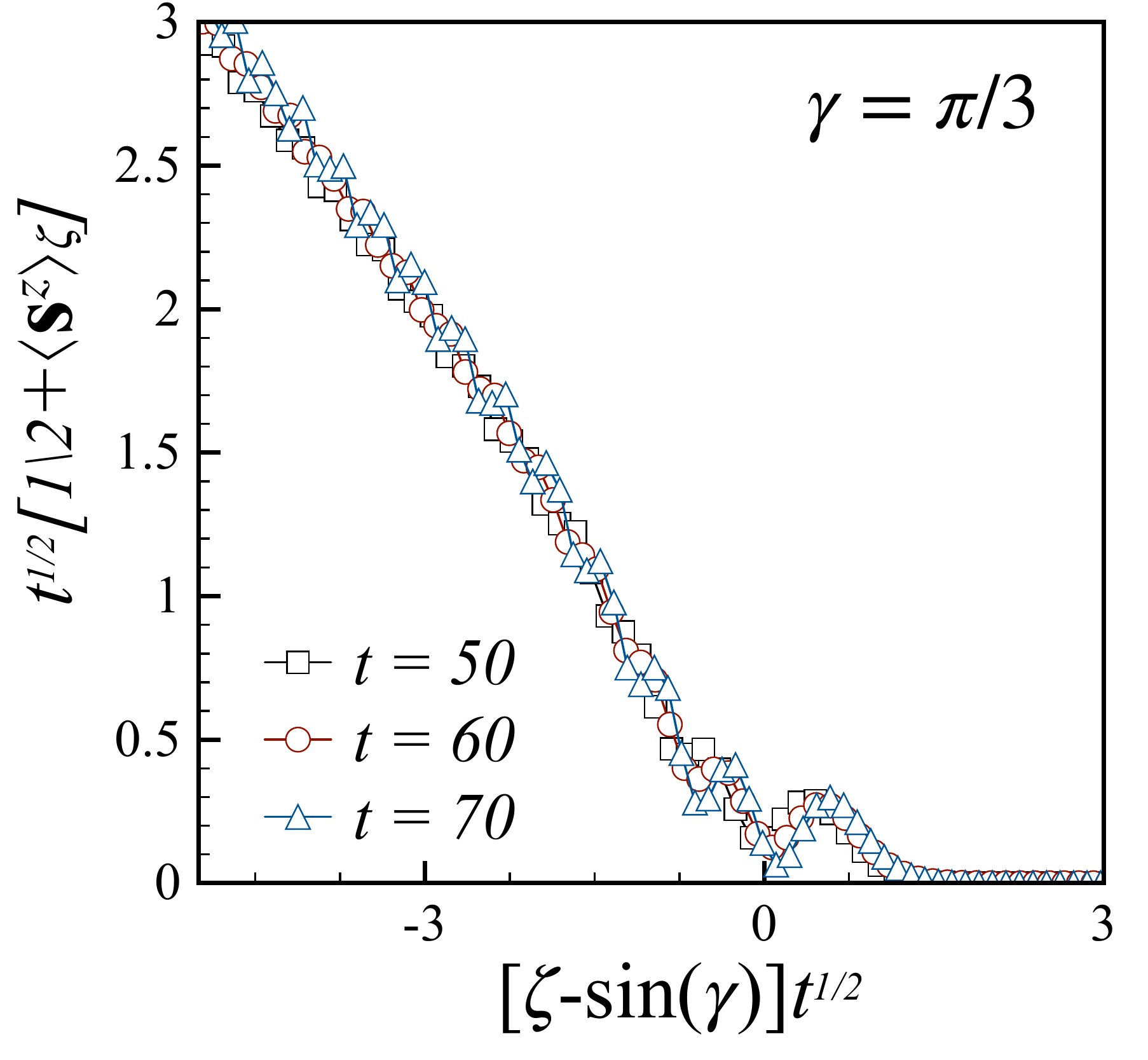}\\
\includegraphics[width=0.24\textwidth]{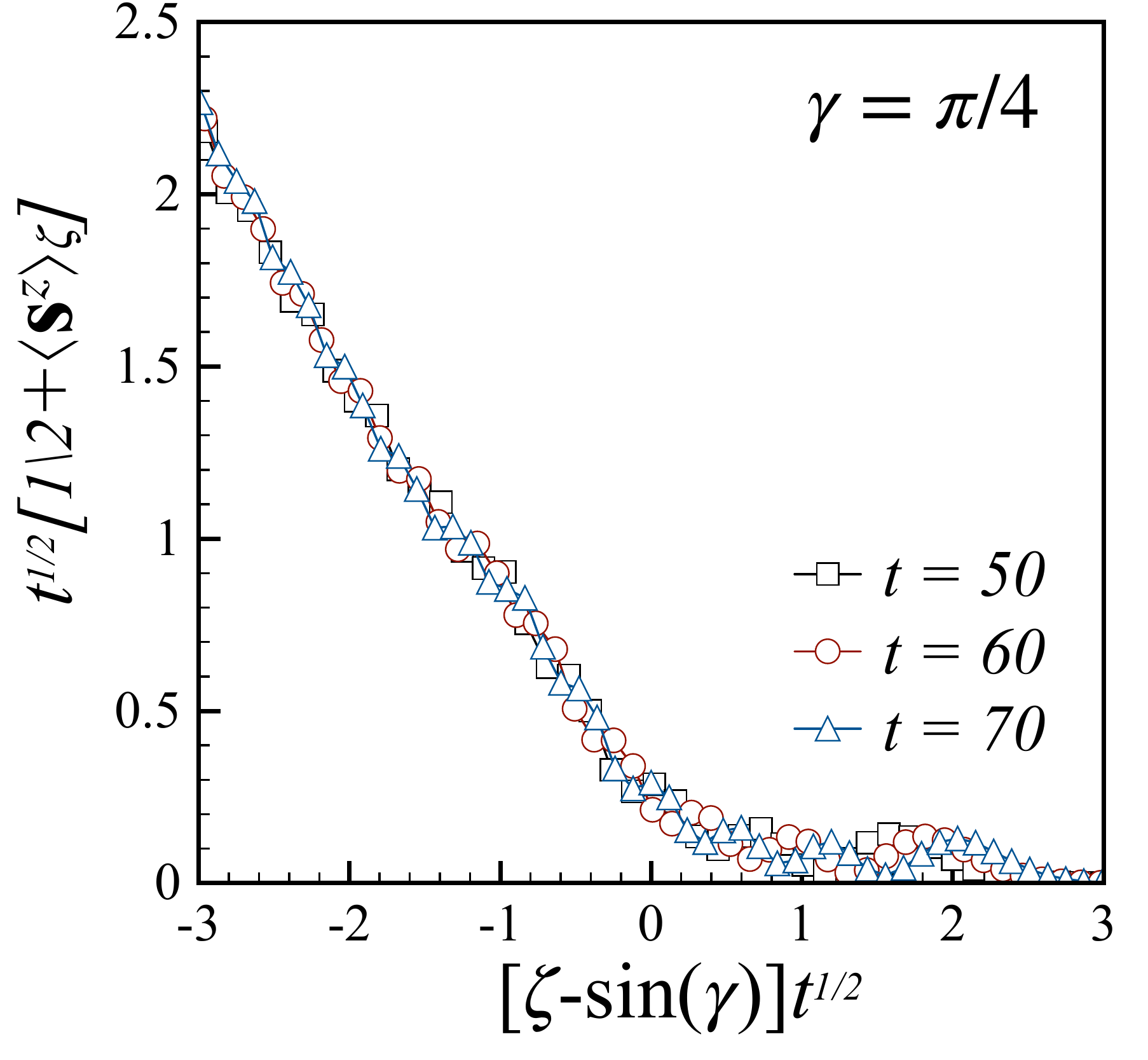}\includegraphics[width=0.24\textwidth]{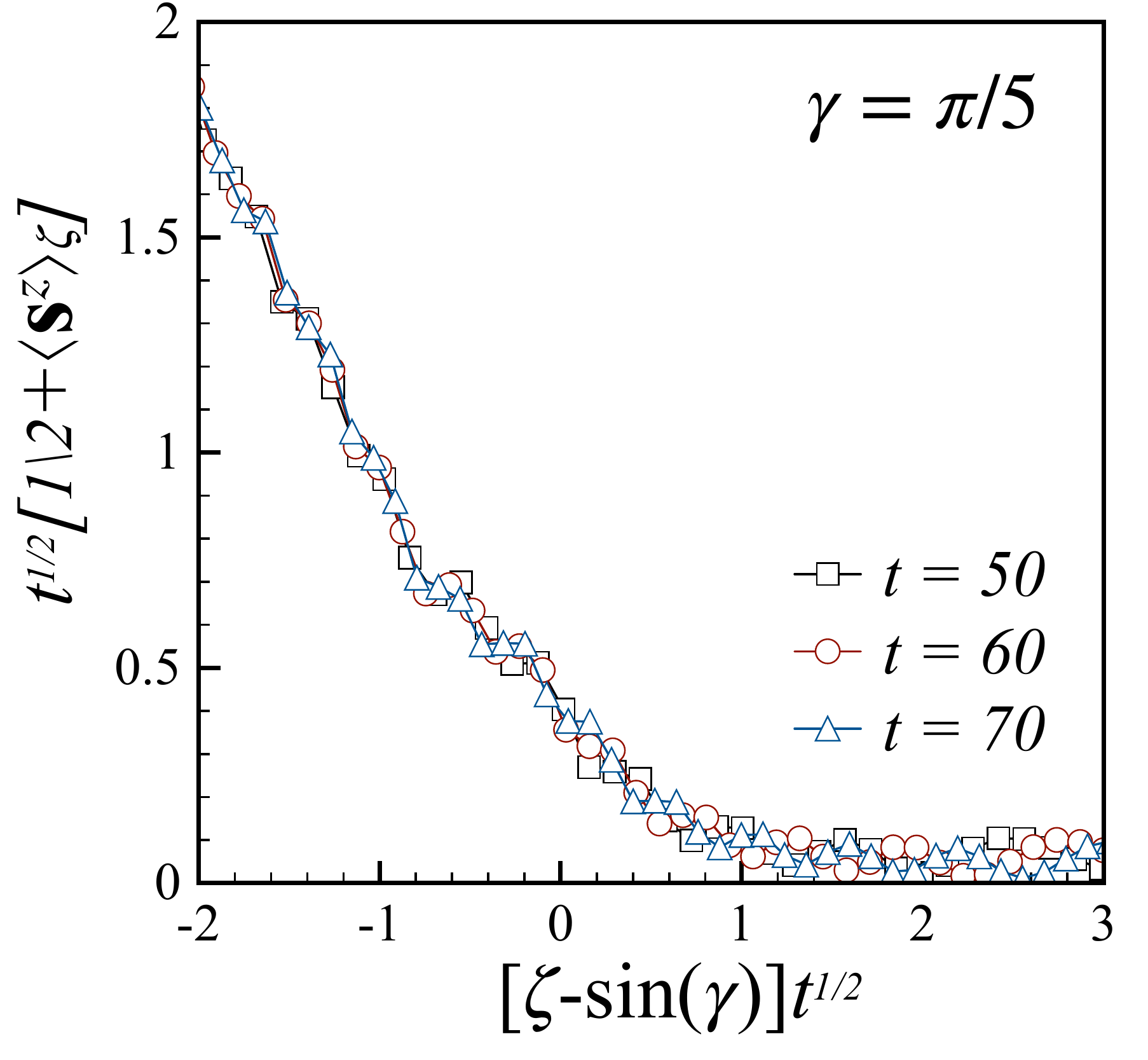}
\caption{\label{fig:edges}
Scaling of the particle density profile at the edge of the light-cone for different values
of the interactions and DW initial condition. As expected, in the free-fermion case
 the scaling of the data is governed by the Airy kernel.
However, when interactions are turned on, the behavior becomes purely diffusive 
preventing a Tracy-Widom like scaling.
}
\end{figure}

\paragraph{Absence of Tracy-Widom distribution and diffusion.---} The profiles in \eqref{eq:profiles} exhibit a smooth dependence on the scaling variable $\zeta$,
apart from the edges of the light-cone, i.e. $\zeta = \pm \sin(\gamma)$, where the derivatives
are non-analytic. In particular, one has 
\be\label{eq:js_der}
\frac{\partial_{\zeta}\langle {\boldsymbol j}_{{\boldsymbol s}^{z}}\rangle_{\zeta=\sin(\gamma)}}{\tanh(h)} = 
-\frac{\tan{(\pi/P)}}{2\pi/P}
\ee
which remains finite for any value of $\gamma$ but $\pi/2$, i.e. the free-fermion point, where it diverges indicating a square root singularity.
The absence of such a singularity in the magnetisation and current profiles for
$|\zeta|=\sin(\gamma)$ is a strong hint that the edges of the front cannot be described by a Tracy-Widom scaling~\cite{TW, ViktorKPZ} as soon as $\Delta\not= 0$ and the model
is interacting. Given the absence of dressing for the DW initial conditions, it is tempting to re-interpret Eqs.~\eqref{eq:profiles} in terms of free fermions. In the simplest case of principal roots of unity, i.e. 
$\gamma = \pi/P$, the magnetisation profile~(\ref{eq:lqss_s}) can be seen as
the density profile $\langle \boldsymbol\rho\rangle_{\zeta}=1/2+\langle {\boldsymbol s}^{z}\rangle_{\zeta}$ in a fictitious free-fermionic lattice model. Such fermions have dispersion relation $\varepsilon(p)=-\cos(p)$ but with the
momentum  $p$ restricted to $[-\gamma, \gamma]$. This restriction is crucial because if $\gamma < \pi/2$, no particles in the initial state
travel at the maximal velocity. Then, an asymptotic analysis of the fermion density near
the edges shows that $t^{1/2}\langle\boldsymbol\rho\rangle_{\zeta}$ is a function of the scaling variable $X=\frac{x\pm t\sin(\gamma)}{\sqrt{t}}$. In the free-fermion problem such a function can be computed exactly in terms of imaginary error functions~\cite{supplmat}, even though checking its validity for the XXZ spin chain requires extremely large simulation times. Nevertheless, the dependence on the scaling variable $X$ of the
magnetisation profiles at the edges is visible in the Fig.~\ref{fig:edges} and rules out for $\Delta\neq0$ the $t^{1/3}$ scaling characteristic of the Tracy-Widom behavior. 
Finally, we observe that, within this picture, in the isotropic limit
$\gamma\rightarrow 0$, i.e. $\Delta\rightarrow 1^-$, the magnetisation profile is expected to be a scaling function of the ratio $\frac{x}{\sqrt{t}}$, for all values of $h$, thus signaling a diffusive behavior~\cite{McCoyDiffusion, GKSS05, StBr11, LowerBoundProsen}. Similar conclusions are suggested by the return probability, indicating diffusive scaling but with slow corrections~\cite{JM}, providing a possible justification for the anomalous scaling observed in \cite{ProsenDMRG}.


\paragraph{Conclusions.---}

We considered the emblematic non-equilibrium protocol generated 
by joining two domains with opposite magnetisation. 
Exploiting the properties of the XXZ spin-$1/2$ chain,
we were able to find a full analytic solution for the LQSS.
We consequently obtained closed expression for both
the magnetisation and spin current stationary profiles.
Interestingly, our analytic results show a strongly discontinuous behavior 
as a function of the interaction $\Delta$, confirming the predictions obtained via the Drude weight.
Moreover, for the DW initial case 
we took advantage of a free-fermion analogy to fully characterise the 
scaling of the stationary profiles at the edges of the light-cone. 
Such analysis has been supported by numerical DMRG simulations and, 
it gave evidence of the absence of a Tracy-Widom scaling a part for
the noninteracting point $\Delta = 0$. 

Our simple solution is a promising framework to derive
a continuous field theory description of the LQSS, thus extending
the results of \cite{VSDH16, SDCV17, DuSC17} in the presence of interactions. An interesting
outcome would be, for instance, the behavior of the entanglement entropy~\cite{AlCa16}.

\nocite{*}


\begin{acknowledgments} 
We are extremely grateful to J. Dubail and J-M. St\'ephan for many stimulating conversations.
M.C. acknowledges support by the Marie Sklodowska-Curie Grant No. 701221 NET4IQ.
This work was supported by  EPSRC Quantum Matter in and out of Equilibrium
Ref. EP/N01930X/1 (A.D.L.).\\

All authors equally contributed to the developing and interpretation 
of theory, results and numerical data,
and to the writing of the manuscript.
\end{acknowledgments}


\onecolumngrid
\newpage
\setcounter{equation}{0}%
\setcounter{figure}{0}%
\setcounter{table}{0}%
\renewcommand{\thetable}{S\arabic{table}}
\renewcommand{\theequation}{S\arabic{equation}}
\renewcommand{\thefigure}{S\arabic{figure}}

\newcounter{app}
\setcounter{app}{0}
\newcounter{app2}
\setcounter{app2}{1}
\newcommand{\appsection}[1]{\setcounter{app2}{1}\stepcounter{app}\section{\Alph{app}.\hspace{1ex}#1}}
\newcommand{\appsubsection}[1]{\subsection{\Alph{app}\arabic{app2}.\hspace{1ex} #1}\stepcounter{app2}}
\begin{center}
{\Large Supplementary Material \\ 
\titleinfo
}
\end{center}
\section{String properties}
Here we summarize the general rule to determine the 
{\it parity} $\pp_j$, {\it lenght} $n_j$ and {\it sign} $\sigma_j$ of a specific string.
Let us recall that we defined $\Delta = \cos(\gamma)$ with $\gamma = \pi \, Q/ P = [0; \nu_1, \nu_2, \dots, \nu_\delta]$.
Following Ref. \onlinecite{takahashi}, let's start by introducing the two series of numbers
$\{y_{-1},y_{0},\dots , y_{\delta}\}$ and $\{m_{0}, m_{1}, \dots, m_{\delta}\}$,
\begin{align}
y_{i} & = \nu_{i} y_{i-1} + y_{i-2}\; , \quad y_{0} = 1 \;,\quad y_{-1} = 0\;, \\
m_{i} & = \sum_{j=1}^{i} \nu_{j} \;, \quad m_{0} = 0\;,
\end{align}
in therms of which we have the following relation for the {\it length} $n_j$:
\be
n_{j} = y_{i-1} + (j-m_{i}) y_{i} \quad {\rm for} \quad m_{i} \le j < m_{i+1} \; ,
\ee
the {\it parity} $\pp_j$:
\be
 \pp_{m_1} = -1 \;,  \quad \pp_{j} = (-1)^{\left\lfloor (n_{j} - 1)\frac{Q}{P} \right\rfloor} \quad  {\rm for} \quad j \neq m_{1} \; ,
\ee
and the {\it sign} $\sigma_j$:
\be
\sigma_{j} = (-1)^{i} \quad {\rm for} \quad m_{i} \le j < m_{i+1} \; .
\ee

Finally, let us collect some useful relations involving the last strings
(where we used the definition $\ell = m_{\delta}$ and the fact that $y_{\delta} = P$):
\begin{align}
n_{\ell} = y_{\delta - 1}\; \quad n_{\ell} + n_{\ell-1} = P \; , \quad  n_{\ell-1} - n_{\ell} = n_{\ell-2} \; , \quad 
\sigma_{\ell-1} = - \sigma_{\ell} \; , \quad \sigma_{\ell} \sin(\pi / P) = \pp_{\ell} \sin (n_{\ell} \gamma) \; .
\end{align}

\section{Filling factors for infinite temperature and finite magnetic field state}
The thermodynamic state 
\be
\varrho(h) = 
\frac{\exp(2 h {\boldsymbol S}^{z})}{Z}
\ee
admits a thermodynamic Bethe ansatz description in therms of the 
following filling factors
\begin{align}
&\vartheta^{(h)}_{j} = \left[ \frac{\sinh(y _{i} h)}{\sinh((n_{j}+y_i) h)} \right]^2  \quad
{\rm for}\quad m_{i} \le j < m_{i+1} \quad {\rm and} \quad j< \ell-1 \;,\\
&\vartheta^{(h)}_{\ell-1}  = \frac{1}{1+\kappa \, e^{h P}}  \;, \quad
\vartheta^{(h)}_{\ell} = \frac{\kappa}{\kappa + e^{h P}}\; , \quad 
\kappa \equiv \frac{\sinh(n_{\ell-1} h ) }{ \sinh(n_{\ell} h) }.
\end{align}

Notice that, in the limit $h\to\infty$ we gets the trivial TBA 
description of the reference state $ \ket{ \Uparrow } \equiv \ket{\uparrow \cdots \uparrow}$, 
which obviously reads
\be
\vartheta^{\ket{ \Uparrow } }_{j} = 0 \;, \quad \forall\; j \in \{1,\dots, \ell\}.
\ee
Otherwise, in the opposite limit $h\to -\infty$, we obtain the representation of the completely full state
 $ \ket{ \Downarrow } \equiv \ket{  \downarrow \cdots \downarrow  }$, 
\be
\vartheta^{ \ket{ \Downarrow } }_{j} = \delta_{j,\ell} +  \delta_{j,\ell-1}  \;, \quad \forall\; j \in \{1,\dots, \ell\}.
\ee


\section{Edge behavior at $\Delta$ root of unity}
\begin{figure}[t!]
\centering
\includegraphics[scale=1]{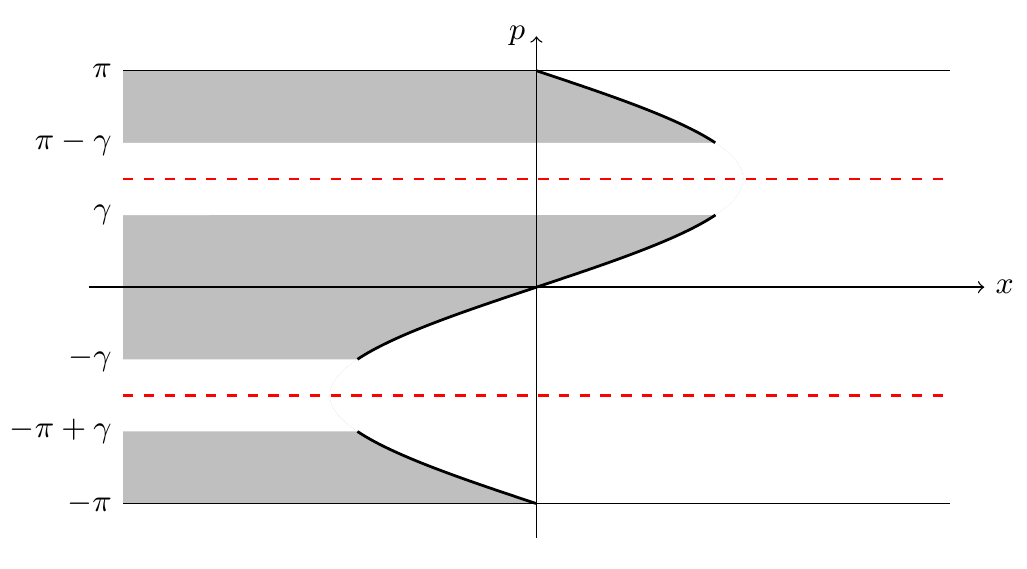}
\vspace*{0cm}~~~\includegraphics[scale=0.7]{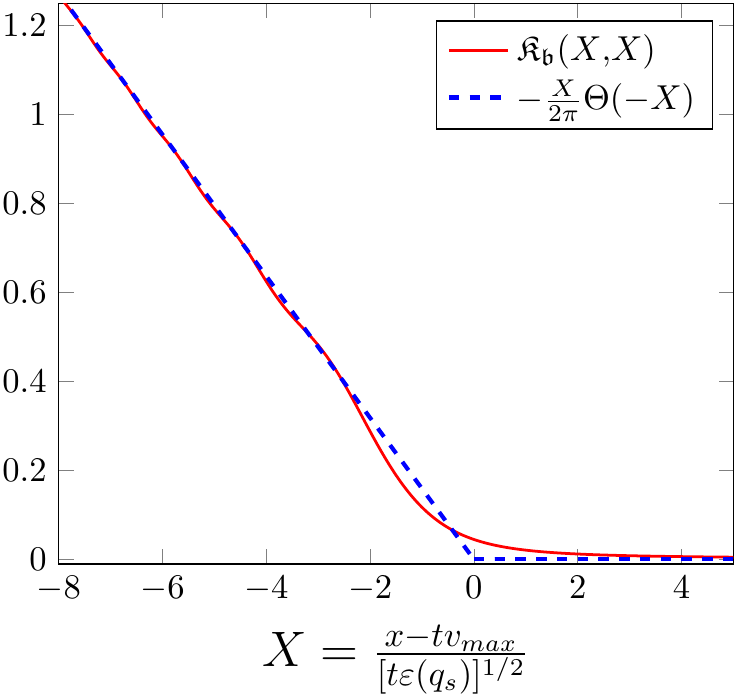}
\caption{\label{fig:hydro}
\textbf{Left.} Hydrodynamic interpretation of the interacting DW quench with anisotropy
$\Delta=\cos(\gamma)$ and $\gamma=\frac{\pi}{\ell}$, $\ell=2,3,\dots$. The last and second-to-last Bethe Ansatz strings are
effectively two non-interacting fermions with momenta that cannot occupy two symmetric intervals centered around
$\pm\pi/2$ and of width $\pi/2-\gamma$. \textbf{Right.} The diagonal part of the error function kernel
$\mathfrak{K}_b$ plotted against the scaling variable $X=\frac{x-t\sin(\gamma)}{[t\cos(\gamma)]^{1/2}}$.
Notice that for large and negative $X$ the error function kernel behaves linearly, alike the magnetisation and current
profiles in the interacting DW quench.
}
\end{figure}

We consider the case $Q=1$ and $P\equiv \ell$, i.e. $\gamma=\pi/\ell$, and the corresponding values of $\Delta$ are called \textit{roots of unity}. Moreover we focus on the limit $h\rightarrow\infty$ that
describes the Domain Wall (DW) initial state. As discussed in the previous section the fillings for the initial states $|\Uparrow\rangle$ and $|\Downarrow\rangle$ are trivial, in particular
$\vartheta^{|\Uparrow\rangle}_{j}(\lambda)=0$  and $\vartheta^{|\Downarrow\rangle}_{j}(\lambda)=\delta_{j,\ell-1}+\delta_{j,\ell}$. The index $j$ takes values $1,\dots,\ell$ according to the
string content when $\Delta$ is a root of unity. The LQSS is described by the fillings
$\vartheta_{j,\zeta}(\lambda)=\vartheta_j^{|\Downarrow\rangle}(\lambda)
\Theta(-v_{j,\zeta}(\lambda)+\zeta)$, being $v_{j,\zeta}$ the dressed velocity~(\ref{eq:v_dressed}). 
Now Eq.~(\ref{eq:dress_q}) in the main text implies that the only non-trivial fillings in the LQSS are the ones of the last two strings. Moreover from the properties of the
kernels it also follows that the momentum and the energy derivatives do not dress for $j=\ell-1,~\ell$, as of course it is implied by the finite-$h$ solution in~(\ref{eq:solution_pprime}). 
The magnetisation profile is given by
  \begin{equation}
  \label{eq:spin_profile}
\langle \boldsymbol s^z\rangle_{\zeta}=\frac{1}{2}-\sum_{j=\ell-1,\ell}n_j\int_{-\infty}^{\infty} d\lambda~\rho_j(\lambda),
  \end{equation}
  that can be rewritten, passing to the bare  momentum variable of the last string $\mathfrak{p}_{\ell}(\lambda)=-2\arctan\bigl[\tan(\gamma/2)\tanh(\lambda)\bigr]$, as
  \begin{equation}
  \label{eq:fermion_spin}
  \langle \boldsymbol s^z\rangle_{\zeta}=-\frac{1}{2}+\frac{1}{2\gamma}\int_{-\gamma}^{\gamma}dp~\Theta(\tilde{v}(p)-\zeta).
  \end{equation}
  It is important to observe  that the integration variable in (\ref{eq:fermion_spin}) is bounded in the interval $p\in[-\gamma,\gamma]$,
  because of the analytic properties of $\mathfrak{p}_{\ell}(\lambda)$ for real $\lambda$. The function $\tilde{v}(p)$ is nothing but the free-fermion
  velocity $\tilde{v}(p)=\sin (p)$. The determination of the magnetisation profile in the DW quench seems formally analogous to the determination of the density profile in a free-fermion problem where two
  strips of width $\pi/2-\gamma$ centered around the the points $p=\pm \pi/2$ are removed from the Brillouin zone. This is illustrated in Fig.~\ref{fig:hydro} on the left. The two allowed bands for the fermions
  correspond to the possible values of the bare momenta for the last and second-to-last string in the Bethe Ansatz solution.
  The contribution of the two bands are identical when calculating the fermion density and we can
  focus only on one of the two.
  
We therefore analyse the DW quench in a free fermion problem with dispersion relation $\varepsilon(p)=-\cos(p)$ and momenta restricted to $[-\gamma,\gamma]$. Notice that we
need however to correctly
normalise the fermion density $\langle\boldsymbol\rho\rangle_{\zeta}$, since in a \textit{bona fide} free-fermion model we would have for $\zeta<-1$,
$\langle\boldsymbol\rho\rangle_{\zeta}=\frac{2\gamma}{\pi}$,
whereas in the DW
quench $\langle\boldsymbol\rho\rangle_{\zeta}=1$ for $\zeta<-1$. Letting aside this issue we consider the fermion propagator~\cite{VSDH16}
\begin{equation}
\label{eq:fermion_prop}
\mathcal{G}_{x,y}(t)=\frac{2\pi}{\gamma}\int_{-\gamma}^{\gamma}\frac{dk}{2\pi}\int_{-\gamma}^{\gamma}\frac{dq}{2\pi}\frac{e^{-i(t \cos k+xk-t\cos q-yq)}}{1-e^{i(q-k+i0)}};
\end{equation}
the density profile obtained from such an integral is $\langle\boldsymbol\rho\rangle_{\zeta}=\langle\boldsymbol s^z\rangle_{\zeta}+1/2$,
being $\langle\boldsymbol s^z\rangle_{\zeta}$ as in \eqref{eq:spin_profile}. The stationary points of the integral satisfy,
for instance in the variable $q$, the equation
$\tilde{v}(q_s)=\zeta$ and therefore the light-cone boundaries are obtained from the condition $\zeta_{\pm}=\pm\max_{q\in[-\gamma,\gamma]}\tilde{v}(q)$.
If $\gamma>\pi/2$ the light-cone
boundaries are at $\zeta_{\pm}=\pm 1$ whereas if $\gamma<\pi/2$ they are located at $\zeta_{\pm}=\pm\sin(\gamma)$. These two cases lead to  different scalings for the fermion propagator
~\eqref{eq:fermion_prop}. Indeed for $\gamma>\pi/2$, the uniform asymptotic in a neighborhood of $\zeta\rightarrow\zeta_{\pm}$ is  obtained by a cubic polynomial approximation of
the phase, due to
the coalescence of two stationary points~\cite{VSDH16}. The result of the stationary phase approximation shows that the fermion propagator is proportional to the Airy kernel~\cite{TW}. However
when $\gamma<\pi/2$, as in the fermion model associated to the interacting DW quench, the change in asymptotic for $\zeta>|\zeta_{\pm}|$ is consequence of a stationary point leaving the
domain of 
integration. As discussed for instance in~\cite{Wong}, a uniform asymptotics is obtained through a quadratic approximation of the phase.  For instance, in a neighborhood of
$\zeta_{+}=\sin(\gamma)$, one gets
\begin{equation}
\label{eq:prop_asympt}
\mathcal{G}_{x,y}(t)=\frac{2\pi e^{-i\gamma(x-y)}}{\gamma[t\cos(\gamma)]^{1/2}}~\mathfrak{K_b}(X,Y)+o(t^{1/2})
\end{equation}
where we defined the error function kernel $\mathfrak{K_b}$
\begin{equation}
 \mathfrak{K_b}(X,Y)=\int_{0}^{\infty}\frac{dK}{2\pi}\int_{0}^{\infty}\frac{dQ}{2\pi}~\frac{e^{iKX+i\frac{K^2}{2}-iQY-i\frac{Q^2}{2}}}{i(Q-K-i0)},
 \end{equation}
 and the scaling variable $X=\frac{x-t\sin(\gamma)}{[t\cos(\gamma)]^{1/2}}$. For convenience we also introduce here the function
 \begin{equation}
U(X)=\int_{0}^{\infty}\frac{dQ}{2\pi}e^{iXQ+i\frac{Q^2}{2}},
\end{equation}
in terms of which the error function kernel satisfies
 $-(\partial_X+\partial_Y)\mathfrak{K_b}(X,Y)=U(X)\overline{U(Y)}$ and $\overline{\mathfrak{K_b}(X,Y)}=\mathfrak{K_b}(Y,X)$. It follows therefore that
 $\mathfrak{K_b}(X,X)$ is real and monotonically decreasing.
 We can determine exactly the diagonal part of the kernel integrating the differential equation $-\frac{d}{dX}\mathfrak{K_b}(X,X)=|U(X)|^2$
 with the boundary condition $\mathfrak{K_b}(X,X)=0$ for $X\rightarrow\infty$; one finds
 \begin{equation}
 \label{density}
 \mathfrak{K_b}(X,X)=-X|U(X)|^2+\frac{\mathrm{Im}[U(X)]}{\pi}.
 \end{equation} 
 Notice that if we expand for  large and negative $X$ the diagonal part of the kernel we obtain the
 asymptotic expansion $\mathfrak{K_b}(X)=-\frac{X}{2\pi}+O(1/X)$, that
 is we recover the expected linear behaviour near the light-cone of the density profile from 
 \eqref{eq:spin_profile}  (see also Fig.~\ref{fig:hydro} on the right)
 \begin{equation}
 \langle\boldsymbol\rho\rangle_{\zeta}\simeq -\frac{1}{\gamma\cos(\gamma)}[\zeta-\sin(\gamma)].
 \end{equation}
 Expanding for large and negative $X$ the Airy kernel we would
 find instead at leading order $\frac{1}{\pi}\sqrt{-X}$; namely a square root singularity in the fermion density
 near the edge of the light-cone. We remind that this case the correct scaling variable is however
 $X=\frac{x-t}{(t/2)^{1/3}}$.



\end{document}